\documentclass[11pt]{article}
\usepackage{latexsym}
\raggedright
\newcommand{\T}{{\rm Tr} \,}

\newcommand{\cB}{{\cal B}}
\newcommand{\cH}{{\cal H}}

\newcommand{\1}{{\bf 1}}

\begin{document}
  \title{Antilinearity in Bipartite Quantum Systems and
 Teleportation}
  \author{Armin Uhlmann}
  \date{Institute
  for Theoretical Physics\\ University of Leipzig}
  \maketitle


In this paper I consider some assorted antilinear operations
and operators
in bipartite quantum systems, an application to quantum
teleportation, and a link to Tomita and Takasaki's theory
via twisted direct products. The idea is in exploring the
natural antilinearity which is inherent to vectors in
direct products of Hilbert spaces. The reason for the
appearance of certain antilinear maps, here called EPR-maps,
is explained in the first section, together with
some basic equations. The acronym EPR stands for the
problem, raised in \cite{EPR35} by Einstein, Podolski,
and Rosen, see also \cite{Peres93}, \cite{NC00}.

Antlinearity in the EPR-problem has been explicitly noticed
by Fivel \cite{Fivel95}.
Here I follow a more general line \cite{Uh00a}, \cite{Uh00b}.
Of course, the exposition in the first section
(and in the third one) are mathematically
near to almost every treatment in which purification and related
topics play there role. Antilinearity is often masked by
introducing distinguished basis in the parts of the bipartite
system.
An interesting different approach
is by Ohya and Belavkin, \cite{BO00}, \cite{Be01}, and
by Ohya's idea of compound states \cite{Oh83}.

In section 2 I present an application to imperfect (unfaithful)
quantum teleportation: Linear teleportations maps
allow for a unique decomposition into pairs of
EPR-maps. Uniqueness would be lost by requiring linearity
due to an ambiguity in phases.

Two norm estimates are derived. The case of
L\"uders measurements with projections of any rank
is considered. An example with distributed measurements
is presented, showing the use of antilinear EPR-maps in
a multipartite system.

The polar decompositions of EPR-maps are considered in
section 3, a rather straightforward task. In these
decompositions the positive parts are the square roots of
the density operators seen in the two subsystems.
The phase operators must be antlinear
partial isometries between the two parts of the direct
Hilbert space product. As explained in section 4, this
feature allows to perform twisted direct products.
They will be compared with an elementary case of well
known operators known from Tomita-Takesaki theory.

In view of
applications to quantum information theory, and
to underline the difference to classical intuition, one often
assumes a macroscopic distance between the two systems.
Though this is reflected in the formalism only rudimentarily,
it provides a nice heuristics: The subsystems can be distinguished
classically, their owners, Alice and Bob, can exchange
classical information (using, say a telephon),
and they are independent one from another.
If they like to perform quantum operations, they
have access just to their parts. Notice that
a macroscopic spatial distance between them is sufficient
for
the observables of Alice to belong to the commutant of
Bob's observables. Of course,
parts of a composed quantum system can be independent
one from another without sitting in spatially different regions.

Remarks on notation: In this paper the Hermitian adjoint of a map
or  of an operator $A$ is denoted by $A^*$. The scalar product
in Hilbert spaces is assumed linear in its second argument.
Sometimes the symbol $\circ$ is used to see more clearly
how maps are composed.

\section{Some basic facts}
Our bipartite quantum systems lives on the
direct product $\cH := \cH_a \otimes \cH_b$ of two Hilbert
spaces, $\cH_a$ and $\cH_b$, with any dimensions.
(A nice little exercise is to follow the formalism in case
of a 1-dimensional $\cH_b$.)
It is a well known fact that
$\cH$ is canonically isomorphic to the space of
Hilbert-Schmidt maps from $\cH_a$ into the dual $\cH_b^*$
of $\cH_b$.
$$
\cH = \cH_a \otimes \cH_b \simeq {\cal L}^2(\cH_a, \cH_b^*)
\simeq {\cal L}^2(\cH_b, \cH_a^*)
$$
$\cH_b^*$ is antilinearly (or conjugate linearly) isomorphic
to $\cH_b$, a fact which is on the heart of Dirac's bra-
ket-formalism $|x \rangle \leftrightarrow \langle x|$.
Composing the bra-ket morphism with the Hilbert-Schmidt maps from
$\cH_a$ into $\cH_b^*$ we get the {\em space of antilinear
Hilbert-Schmidt maps} from $\cH_a$ into $\cH_b$. Indicating
the antilinearity by an index {\em anti}, we
have the natural isomorphisms
\begin{equation} \label{1}
\cH_a \otimes \cH_b \simeq {\cal L}^2_{\rm anti}(\cH_a, \cH_b)
\simeq {\cal L}^2_{\rm anti}(\cH_b, \cH_a)
\end{equation}
Let us look at these morphisms in more detail, and let us start
with an arbitrary vector $\psi$ from $\cH$. There are
decompositions
\begin{equation} \label{2}
\psi = \sum \phi_k^a \otimes \phi_k^b, \quad
\phi_i^a \in \cH_a, \, \phi_i^b \in \cH_b
\end{equation}
converging in norm. Choosing one of them arbitrarily, we set
\begin{equation} \label{3}
{\bf s}_{\psi}^{ba} \phi^a := \sum \langle \phi^a, \phi_k^a \rangle
\, \phi_k^b
\end{equation}
Every member of the sum is a map from $\cH_a$ into $\cH_b$.
Their 2-norms are the same as the norm of the corresponding term
in the decomposition (\ref{2}). Hence, (\ref{3}) defines an
antilinear Hilbert-Schmidt
map from $\cH_a$ into $\cH_b$. Its adjoint, a map
from $\cH_b$ into $\cH_a$, is defined by the relation
\begin{equation} \label{4}
\langle \phi^b, {\bf s}_{\psi}^{ba} \phi^a \rangle =
\langle \phi^a, ({\bf s}_{\psi}^{ba})^* \phi^b \rangle
\end{equation}
for all $\phi^a$ and $\phi^b$. By an evident calculation one
gets
\begin{equation} \label{5}
({\bf s}_{\psi}^{ba})^* \phi^b =
\sum \langle \phi^b, \phi_k^b \rangle \, \phi_k^a
\end{equation}
and we denote this map in accordance with (\ref{3}) by
${\bf s}_{\psi}^{ab}$.

In the next step we explicitly see the independence of the
constructions from the chosen decomposition (\ref{2}) of $\psi$.
It provides the contact to a famous problem of Einstein, Rosen,
and Podolski \cite{EPR35}. Assume the state of the bipartite
system is defined by $\psi \in \cH$.
If Alice does a measurement with one
of her observables, $A \in \cB(\cH_a)$, her activity is a
measurement in {\em every} larger quantum system
which contains Alice's system. In particular, this is the case
in the bipartite system based on $\cH$. Here the relevant
observable reads $A \otimes \1^b$.

We now choose
Alice's observable to be the rank one projection
$P = |\phi^a \rangle\langle \phi^a|$, $\phi^a \in \cH_a$
being a unit vector. In doing so, the
measurement terminates in showing randomly the eigenvalue 1
or 0 of $P$. In case it shows the eigenvalue 1, the state vector
of the bipartite system system has switched from $\psi$
to $(P \otimes \1^b) \psi$. A new state vector
has been prepared.

Our aim, to show the independence of (\ref{3}) from the
chosen decomposition (\ref{2}) of $\psi$, is reached by
proving
\begin{equation} \label{6}
(|\phi^a  \rangle\langle \phi^a| \otimes \1^b) \psi
= \phi^a \otimes {\bf s}_{\psi}^{ba} \phi^a, \quad
\forall \phi^a \in \cH_a
\end{equation}
To show (6) for a given decomposition of $\psi$,
one first remarks the linear dependence of (\ref{3}) from
the terms of the sum (\ref{2}). Thus, one has to check
(\ref{6}) just for product vectors, a simple task.
Remark that a similar relation holds for an appropriate
action of Bob.

In conclusion we have seen that every $\psi \in \cH$
uniquely determines antilinear Hilbert-Schmidt maps
according to (\ref{3}) and (\ref{5}).  Let us call them
the {\em EPR-maps} belonging to $\psi$. They are
antilinear equivalents of $\psi$ obeying
\begin{equation} \label{7}
({\bf s}_{\psi}^{ba})^* = {\bf s}_{\psi}^{ab},
\quad
({\bf s}_{\psi}^{ab})^* = {\bf s}_{\psi}^{ba}
\end{equation}
In rewriting (\ref{4}) and (\ref{7}), we can add a
conclusion seen from (\ref{6}): It holds
\begin{equation} \label{8}
\langle \phi^b , {\bf s}_{\psi}^{ba} \phi^a \rangle =
\langle \phi^a , {\bf s}_{\psi}^{ab} \phi^b \rangle =
\langle \phi^a \otimes \phi^b , \psi \rangle
\end{equation}
for all $\phi^a \in \cH_a$, $\phi^b \in \cH_b$, and
$\psi \in \cH$. Now we proceed as follows: Because
every $\varphi \in \cH$ can be written as a sum of
product vectors, we try to calculate its scalar
product with $\psi$ by the help of (\ref{8}). A more
or less straightforward calculation will show the
validity of
\begin{equation} \label{9}
\langle \varphi , \psi \rangle =
{\rm Tr}_a {\bf s}_{\psi}^{ab} {\bf s}_{\varphi}^{ba} =
{\rm Tr}_b {\bf s}_{\psi}^{ba} {\bf s}_{\varphi}^{ab}
\end{equation}
the right hand term of which are, in view of (\ref{7}),
antilinear versions of the von Neumann scalar product.
Let me add that one can derive (\ref{8}) from
(\ref{9}) by choosing $\varphi = \phi^a \otimes \phi^b$.

What remains for a first account is the reconstruction of
$\psi$ from one of its EPR-maps.  The task
can be done with the help of any decomposition of the unit operator
$\1^a$ of, say, Alice. More generally, let $A \in \cB(\cH_a)$
be a positive operator and
\begin{equation} \label{10}
A = \sum | \phi_k^a \rangle\langle \phi_k^a |
\end{equation}
a rank one decomposition of $A$. Then
\begin{equation} \label{11}
A \, \psi = \sum \phi_k^a \otimes {\bf s}_{\psi}^{ba} \phi_k^a
\end{equation}
$\psi$ is returned with Alice's unit operator, $A = \1^a$.

The reduced density operator, $\omega^a_{\psi}$,
can be defined by
$$
{\rm Tr}_a A \omega^a_{\psi} =
\langle \psi, (A \otimes \1^b) \psi \rangle,
\quad A \in \cB(\cH_a)
$$
Similar one gets $\omega^b_{\psi}$ by letting play Bob the role
of Alice.  What one can learn from (\ref{6}) and (\ref{7}) is
\begin{equation} \label{12}
\omega^a_{\psi} = {\bf s}_{\psi}^{ab} {\bf s}_{\psi}^{ba},
\quad
\omega^b_{\psi} = {\bf s}_{\psi}^{ba} {\bf s}_{\psi}^{ab}
\end{equation}

Finally we consider two vectors which are related by
\begin{equation} \label{e3}
\varphi = (A \otimes B) \psi
\end{equation}
In terms of EPR-maps the relation converts to
\begin{equation} \label{e4}
{\bf s}_{\varphi}^{ba} = B {\bf s}_{\psi}^{ba} A^*,
\quad
{\bf s}_{\varphi}^{ab} = A {\bf s}_{\psi}^{ab} B^*
\end{equation}

\section{Imperfect quantum teleportation}

In \cite{BBCJPW93} Bennett {\it et al} invented a protocol,
the BBCJPW-protocol, allowing
for faithful teleportation of vectors and of general states between
Hilbert spaces of finite and equal dimensions $d$. It consists
of one classical information channel and $d^2$ quantum channels.
The latter are randomly triggered by a Bell-like
von Neumann measurement.
The information, which quantum channel has been activated,
is carried by the classical channel. It serves to reconstruct,
by a unitary move, the desired state at the destination.
The protocol has been programmed as a quantum circuit by
Brassard \cite{BBC97}.

A general and self-consistent discussion of all perfect
teleportation schemes and their relation to dense coding
has been given recently by Werner \cite{We00}.

All these tasks and protocols need reference frames (computational
basis) in order to define whether the original
and the teleported vectors (or general states) should be
considered as equal ones or not.
Notice: the  problem is not to tell which of the quantum
channels is triggered nor to identify its output. It is the question
how to relate the input to the output. Usually the problem is
solved by  {\em given} reference basis, one in the
input and one in the output space.
Every reference base determines a conjugation. These conjugations,
composed with the canonical antilinear maps, mask the natural
antilinearity in all these protocols.

Now I am going to describe the way antilinearity enters in
the handling of general, possibly imperfect, unfaithful
teleportation channels.
Let $\cH$ be a tripartite Hilbert space
\begin{equation} \label{13}
 \cH_{abc} = \cH_a \otimes \cH_b \otimes \cH_c
\end{equation}
The {\em input} is an {\em unknown} vector $\phi^a \in \cH_a$.
One further needs a {\em resource} which provides
the so-called {\em entanglement}  \cite{Schr35a} between
the $b$- and the
$c$-system. The resource is given by an {\em ancilla},
mathematically just a {\em known} vector $\varphi^{bc}$,
chosen from $\cH_b \otimes \cH_c$. (More involved, but also
tractable, is the case of an ancilla in a mixed state.)
Thus, the teleportation protocol starts with a vector
\begin{equation} \label{14}
\varphi^{abc} := \phi^a \otimes \varphi^{bc} \in \cH_{abc}
\end{equation}
It is triggered by a measurement within the
$ab$-system. We need a measurement which is also preparing.
There should exist an apparatus doing it. But a single apparatus
can only distinguish between finitely many values. The
conclusion is: We have to trigger the protocol by
measuring an observable,
\begin{equation} \label{obs}
A = \sum_{j=1}^m a_j P_j, \quad \sum P_j = \1^{ab},
\end{equation}
in the $ab$-system
which is a {\em finite} sum with mutually different values $a_j$.
The $P_j$ are projection operators, orthogonal one to another,
and decomposing the unit operator of $\cH_{ab}$.
The measurement itself selects randomly one of these projectors
with a well defined probability. If this projection is $P_j$,
then the measuring device points onto the value $a_j$, thus
indicating {\em which} projection is preparing the new state.
The duty of the classical channel is to inform the owner
of the $c$-system which projection has been processing.

For the discussion of the preparing we assume that $P = P^{ab}$
is one of the projectors $P_j$ appearing in (\ref{obs}).
A measurement in the $ab$-subsystem is simultaneously a
measurement in the larger $abc$-system, and there the projection
operator reads $P \otimes \1^c$. Thus, the preparing becomes
\begin{equation} \label{prep}
\phi^a \otimes \varphi^{bc} \, \longrightarrow \,
( P \otimes \1^c) ( \phi^a \otimes \varphi^{bc} )
\end{equation}
We now impose a restrictive assumption in (\ref{prep}):
$P$ should be of rank
one. Thus, $P$ has to test whether the $ab$-system is in a certain
vector state, say $\psi = \psi^{ab}$, or not.
As the main merit of
the assumption, the prepared state gets the special form
\begin{equation} \label{15}
( |\psi^{ab} \rangle\langle \psi^{ab}| \otimes \1^c)
( \phi^a \otimes \varphi^{bc} )
= \psi^{ab} \otimes \phi^c,
\end{equation}
determining $\phi^c \in \cH_c$.
Varying $\phi^a$ we now define the
map ${\bf t}^{ca}_{\psi, \varphi}$ by
\begin{equation} \label{17}
 {\bf t}^{ca}_{\psi, \varphi} \phi^a = \phi^c
\end{equation}
The {\em teleportation map ${\bf t}^{ca}_{\psi, \varphi}$},
or ${\bf t}^{ca}$ for short, can be computed, \cite{Uh00a},
by
\begin{equation} \label{18}
{\bf t}^{ca}_{\psi, \varphi} = {\bf s}_{\varphi}^{cb} \circ
{\bf s}_{\psi}^{ba}
\end{equation}
This is the {\em factorization property}, valid for every (imperfect)
teleportation channel under the condition that the preparing
projection operator is of rank one. There is no
restriction otherwise, neither on the dimensions of
the Hilbert spaces,
nor on the ancillary vector $\varphi$ or on the vector $\psi$.

The proof is mainly an exercise in
algebraic manipulations, while the convergence problems are
rather harmless due to the Hilbert-Schmidt property of the
two maps involved. With a basis $\phi_1^b, \phi_2^b, \dots$,
of $\cH_b$ we write, according to (\ref{11})
$$
\varphi^{abc} = \sum \phi^a \otimes \phi_j^b \otimes
{\bf s}_{\varphi}^{cb} \phi^a
$$
Next, this expression inserted into (\ref{15}) yields
$$
\psi^{ab} \otimes \phi^c = \sum |\psi^{ab} \rangle\langle
\psi^{ab}, \phi^a \otimes \phi_j^b \rangle \otimes
{\bf s}_{\varphi}^{cb} \phi_j^b
$$
(\ref{8}) allows to rewrite the scalar product to get
$$
\psi^{ab} \otimes \phi^c = \sum
\langle {\bf s}_{\psi}^{ba} \phi^a, \phi^b_j \rangle
\psi^{ab} \otimes {\bf s}_{\varphi}^{cb} \phi_j^b
$$
The antilinearity of the EPR-map converts the right hand
side into
$$
\sum \psi^{ab} \otimes {\bf s}_{\varphi}^{cb}
\langle \phi^b_j \rangle, {\bf s}_{\psi}^{ba} \phi^a \rangle
\phi_j^b =
\psi^{ab} \otimes {\bf s}_{\varphi}^{cb} \circ
{\bf s}_{\psi}^{ba} \phi^a
$$
which is the assertion.

Before looking at some applications of the factorization theorem,
I mention that Alberio and Fei, \cite{AF00}, derived  a condition
for a generally imperfect channel to become faithful.

\subsection{Estimates}
The high symmetry provided by maximally entangled vector
states used in faithful teleportation schemes,
\cite{BBCJPW93}, \cite{We00}, is broken in imperfect
teleportation. As a result, some of the vectors in $\cH_a$
are more efficiently transported than others. Therefore,
the highest possible transport probability is of some
interest.

Let $\phi^a$, $\psi$, $\varphi$ be unit vectors. The probability
for the process $\phi^a \to \phi^c$ is
$$
\langle \phi^c , \phi^c \rangle =
\langle {\bf t}^{ca} \phi^a , {\bf t}^{ca} \phi^a \rangle
$$
Because $\psi$ and $\varphi$ are vectors of two bipartite
systems, and $\cH_b$ is a part of both systems, we can
compare their reductions to the b-system. Call $\varrho$
and $\omega$ the two reduced density operators living
in the b-system. One can prove, see (\ref{2.1}) and (\ref{2.2})
below,
\begin{equation} \label{19}
\langle \phi^c , \phi^c \rangle \leq
| \sqrt{\omega} \varrho \sqrt{\omega} |_{\infty}
\end{equation}
for all unit vectors in $\phi^c \in \cH^c$.
The norm used at the right hand side is the operator norm.
The norm
of a positive operator is its largest eigenvalue.

Being of trace class, one would like to estimate the effectivity
of the single teleportation map by the trace norm.
Interesting enough, the trace norm of ${\bf t}^{ca}$ is the
square root of the transition probability (or fidelity)
between $\varrho$ and $\omega$,
\begin{equation} \label{20}
| {\bf t}^{ca} |_1 = F(\varrho, \omega) =
\T (\sqrt{\omega} \varrho \sqrt{\omega} )^{1/2}
\end{equation}
The estimates are in line with the question how to optimize
quantum teleportation. Depending on specific demands, the problem
has been addressed by
Horodecki {\it et al} \cite{HHH96}, Trump {\it et al} \cite{TBL01},
Banaczek \cite{Ba01}, \v Reh\`a\v cek {\it et al} \cite{RHFB01}.

\subsection{L\"uders measurements}
It is a strong assumption, to suppose Alice could perform rank
one measurements. With raising magnitude of degrees of
freedom the task become more and more difficult.
 In the realm of relativistic quantum field theories
local measurements with projections of infinite rank
are most natural. (Though these systems contain lots of
finite dimensional subsystems, one has to find some with
sufficiently exposed sets of quantum levels.)
Thus, the projection $P$ in the preparing
step (\ref{prep}) may be of any rank. Let
\begin{equation} \label{prep2}
P = \sum | \psi_k^{ab} \rangle\langle \psi_k^{ab} |
\end{equation}
be an orthogonal decomposition of $P$ into rank one projection
operators. Associating EPR-maps
\begin{equation} \label{prep3}
\psi_k^{ab} \, \longleftrightarrow \, {\bf s}_k^{ba}
\end{equation}
to every vector appearing in (\ref{prep2}), (\ref{prep})
becomes
\begin{equation} \label{prep4}
( P \otimes \1^c) ( \phi^a \otimes \varphi^{bc} ) =
\sum \psi^{ab}_k \otimes {\bf t}^{ca}_k \phi^a, \quad
{\bf t}^{ca}_k = {\bf s}^{cb}_{\varphi} \circ {\bf s}^{ba}_k
\end{equation}
We have to decouple the degrees of freedom coming from the
$b$-system. To do so, we first convert the maps between
vectors in those between (not necessarily normalized)
density operators. Then we reduce the right hand side of
(\ref{prep4}) to the $c$ system. Abbreviating
$({\bf t}^{ca})^*$ by ${\bf t}^{ac}$,
the result is the map
\begin{equation} \label{prep5}
| \phi^a \rangle\langle \phi^a | \, \longrightarrow \,
\sum {\bf t}^{ca}_k  (| \phi^a \rangle\langle \phi^a |)
{\bf t}^{ac}_k
\end{equation}
We estimate (\ref{prep4}): The norm of the left is smaller
than product of the norms off $\phi^a$ and $\varphi^{bc}$.
On the right side orthogonality of the $\psi_k$ allows
to calculate the norm. We get
$$
\parallel \phi^a \parallel \cdot \parallel \varphi^{bc} \parallel
\geq \bigl(\sum \langle \phi^a, {\bf t}^{ac}_k {\bf t}^{ca}_k
\phi^a \bigr)^{1/2}
$$
Being valid for all vectors from $\cH_a$ we conclude
\begin{equation} \label{prep6}
| \sum {\bf t}^{ac}_k {\bf t}^{ca}_k |_{\infty} \leq
\parallel \varphi^{bc} \parallel
\end{equation}
The boundedness of the operator allows to extend (\ref{prep5})
to a map from the trace class operators on $\cH_a$ to those
of $\cH^c$. The extension reads
\begin{equation} \label{prep7}
{\bf T}^{ca}(\nu^a) := {\bf s}^{cb}_{\varphi}
\bigl( \sum {\bf s}^{ba}_k \nu^a {\bf s}^{ab}_k \bigr)
{\bf s}^{bc}_{\varphi}
\end{equation}
with $\nu^a$ an arbitrary trace class operator. Estimating
the trace of ${\bf T}$ by (\ref{prep6}) one sees
\begin{equation} \label{prep8}
| {\bf T}^{ca} |_1 \leq
\langle \varphi^{bc}, \varphi^{bc} \rangle
\end{equation}
More general, positive operator valued measurements have been
examined by Mor and Horodecki, \cite{MH99}.

\subsection{Distributed measurements}
In a multipartite system with an even number of subsystems
one can distribute the measurements and the entanglement
resources over some pairs of subsystems. Let us see
this with
five subsystems,
\begin{equation} \label{5sub}
\cH = \cH_a \otimes \cH_b \otimes \cH_c \otimes
\cH_d \otimes \cH_e
\end{equation}
The input is an unknown vector $\phi^a \in \cH_a$, the
ancillarian vectors are selected from the $bc$- and the
$de$-system,
$$
\varphi^{bc} \in \cH_{bc}, \quad \varphi^{de} \in \cH_{de}
$$
and the vector of the total system we are starting with is
\begin{equation} \label{5suba}
\varphi \equiv \varphi^{abcde} = \phi^a \otimes
\varphi^{bc} \otimes \varphi^{de}
\end{equation}
The channel is triggered by measurements in the $ab$- and
in the $cd$-system. To see what is going on it suffices
to treat rank one measurements.
Suppose these measurements prepare, if successful,  the
vectors
$$
\psi^{ab} \in \cH_{ab}, \quad \psi^{cd} \in \cH_{cd}
$$
The we get the relation
\begin{equation} \label{5subr}
( |\psi^{ab} \rangle\langle \psi^{ab}| \otimes
 |\psi^{cd} \rangle\langle \psi^{cd}| \otimes \1^e ) \psi
= \psi^{ab} \otimes \psi^{cd} \otimes \phi^e
\end{equation}
and the vector $\phi^a$ is mapped onto $\phi^e = {\bf t}^{ea} \phi^a$.
Introducing the EPR-maps corresponding to the used vectors
$$
\psi^{ab} \to {\bf s}^{ba}, \quad
\varphi^{bc} \to {\bf s}^{cb}, \quad
\psi^{cd} \to {\bf s}^{dc}, \dots,
$$
the {\em factorization property} becomes
\begin{equation} \label{5subf}
{\bf t}^{ea} = {\bf s}^{ed} \circ {\bf s}^{dc}
\circ {\bf s}^{cb} \circ {\bf s}^{ba}
\end{equation}

\section{Polar decompositions}
Coming back to the bipartite case $\psi \in \cH_a \otimes \cH_b$,
we shall explore the polar decompositions of the EPR-maps
${\bf s}_{\psi}^{ba}$ and ${\bf s}_{\psi}^{ab}$.

As we already know by  (\ref{12}), the positive factors in
the polar decompositions
must be the square roots of the reduced density operators,
$\omega^a$ and $\omega^b$, of $\psi$. Their phase operators
are antiunitary partial isometries between the two parts of
the bipartite Hilbert space. We call these maps
${\bf j}_{\psi}^{ba}$ and ${\bf j}_{\psi}^{ab}$.
The first of these antilinear operations maps $\cH_a$ into $\cH_b$,
the second $\cH_b$ into $\cH_a$.
Standard technique yields the {\em polar decompositions}
\begin{equation} \label{p1}
{\bf s}_{\psi}^{ba} = (\omega^b)^{1/2} {\bf j}_{\psi}^{ba}
= {\bf j}_{\psi}^{ba} (\omega^a)^{1/2},
\end{equation}
$$
{\bf s}_{\psi}^{ab} = (\omega^a)^{1/2} {\bf j}_{\psi}^{ab}
= {\bf j}_{\psi}^{ab} (\omega^b)^{1/2}
$$
Just as in the linear case, one requires
\begin{equation} \label{j2}
{\bf j}_{\psi}^{ab}  {\bf j}_{\psi}^{ba} = Q^a, \quad
{\bf j}_{\psi}^{ba}  {\bf j}_{\psi}^{ab} = Q^b
\end{equation}
where $Q^a$, respectively $Q^b$, is the projection operator
onto the support space of $\omega^a$, respectively of
$\omega^b$. The unicity of the polar decomposition and
(\ref{7}) yield
\begin{equation} \label{j1}
( {\bf j}_{\psi}^{ba} )^* = {\bf j}_{\psi}^{ab}, \quad
\omega^b = {\bf j}_{\psi}^{ba} \omega^a {\bf j}_{\psi}^{ab}
\end{equation}

One can relate the expectation values of the reduced density
operators. Let us prove it as an exercise in antilinearity.
We choose $A \in \cB(\cH_a)$ and $B \in \cB(\cH_b)$ such that
\begin{equation} \label{e1}
B^* {\bf j}_{\psi}^{ba} = {\bf j}_{\psi}^{ba} A
\end{equation}
Then
$$
\T \omega^a A = \T \omega^a {\bf j}_{\psi}^{ab} {\bf j}_{\psi}^{ba}
A = \T \omega^a {\bf j}_{\psi}^{ab} B^* {\bf j}_{\psi}^{ba}
$$
The trace of the products two antilinear operators,
$\vartheta_1 \vartheta_2$, is conjugate complex to
the trace of $\vartheta_2 \vartheta_1$. Hence, the expression
under consideration is the complex conjugate of
$$
\T {\bf j}_{\psi}^{ba} \omega^a {\bf j}_{\psi}^{ab} B^*
= \T \omega^b B^*
$$
In conclusion it follows
\begin{equation} \label{e2}
\T \omega^a A = \T \omega^b B
\end{equation}
from (\ref{e1}).

Another useful observation: Let $\cH'_a \subseteq \cH_a$ be the
supporting subspace of a given density operator $\omega^a$.
The set of  all purifications $\psi$ of $\omega^a$
is in one-to-one correspondence to the set of antilinear isometries
from $\cH'_a$ into $\cH_b$.

Finely, we have a look at some facts from which
the norm estimates of the teleporting maps will follow.
To this end we consider two arbitrary vectors,
$\varphi$ and $\psi$, from $\cH = \cH_a \otimes \cH_b$
with reduced (not normalized) density operators
$\varrho^a$ and $\omega^a$ respectively.
Their polar decompositions, (\ref{p1}), yield
\begin{equation} \label{2.1}
{\bf s}_{\varphi}^{ba} {\bf s}_{\psi}^{ab} =
{\bf j}_{\varphi}^{ba} \sqrt{\varrho^a} \sqrt{\omega^a}
{\bf j}_{\phi}^{ab}
\end{equation}
Therefore, the singular values of the operators
\begin{equation} \label{2.2}
{\bf s}_{\varphi}^{ba} {\bf s}_{\psi}^{ab}, \quad
{\bf s}_{\varphi}^{ab} {\bf s}_{\psi}^{ba}, \quad
(\sqrt{\varrho^a} \omega \sqrt{\varrho^a} )^{1/2}
\end{equation}
are equal one to another. The singular values of a
Hilbert-Schmidt operator $\xi$ are the eigenvalues of the
square root of $\xi^* \xi$. That way one proves (\ref{20})
and similarly, (\ref{19}). Notice that for all
$B \in \cB(\cH_b)$
\begin{equation} \label{2.3}
\T {\bf s}_{\varphi}^{ba} {\bf s}_{\psi}^{ab} B =
\langle \psi, (\1^a \otimes B) \varphi \rangle =
\T (\sqrt{\varrho^a} \omega \sqrt{\varrho^a} )^{1/2}
({\bf j}_{\psi}^{ab} B^* {\bf j}_{\varphi}^{ba})
\end{equation}
As an application let us prove a key statement of
the important
paper on the mixed state cloning problem by Barnum
{\it et al.}\cite{BCFJS96} It asserts
\begin{equation} \label{clone1}
F(\omega_{\psi}^a, \omega_{\varphi}^a) =
F(\omega_{\psi}^b, \omega_{\varphi}^b) \, \longmapsto \,
\omega_{\psi}^a \omega_{\varphi}^a =
\omega_{\varphi}^a \omega_{\psi}^a
\end{equation}
(See (\ref{20}) for the definition of $F$.)
It is well know, and easily derived from (\ref{2.3}),
that the assumption of (\ref{clone1}) is satisfied if and
only if
\begin{equation} \label{clone2}
{\bf s}_{\varphi}^{ba} {\bf s}_{\psi}^{ab} \geq 0,
\quad
{\bf s}_{\varphi}^{ab} {\bf s}_{\psi}^{ba} \geq 0
\end{equation}
To say something new, we shall weaken this assumption in
requiring only hermiticity instead of positivity.
By (\ref{7}) it means
\begin{equation} \label{clone3}
{\bf s}_{\varphi}^{ba} {\bf s}_{\psi}^{ab} =
{\bf s}_{\psi}^{ba} {\bf s}_{\varphi}^{ab},
\quad
{\bf s}_{\varphi}^{ab} {\bf s}_{\psi}^{ba} =
{\bf s}_{\psi}^{ab}  {\bf s}_{\varphi}^{ba}
\end{equation}
In the following, starting with (\ref{12}), we systematically
reorder the appearing factors by the the help of
(\ref{clone3}):
$$
\omega_{\psi}^a \omega_{\varphi}^a =
{\bf s}_{\psi}^{ab} {\bf s}_{\psi}^{ab}
{\bf s}_{\varphi}^{ab} {\bf s}_{\varphi}^{ba} =
{\bf s}_{\psi}^{ab} {\bf s}_{\varphi}^{ba}
{\bf s}_{\psi}^{ab} {\bf s}_{\varphi}^{ba}
$$
$$
{\bf s}_{\psi}^{ab} {\bf s}_{\varphi}^{ba}
{\bf s}_{\psi}^{ab} {\bf s}_{\varphi}^{ba} =
{\bf s}_{\varphi}^{ab} {\bf s}_{\psi}^{ba}
{\bf s}_{\varphi}^{ab} {\bf s}_{\psi}^{ba} =
{\bf s}_{\varphi}^{ab} {\bf s}_{\varphi}^{ba}
{\bf s}_{\psi}^{ab} {\bf s}_{\psi}^{ba}
$$
and, again by (\ref{12}), we are done.

\section{From vectors to Operators on $\cH_a \otimes \cH_b$}
With one or two vectors, drawn from the Hilbert space $\cH$
of our bipartite system, one can associate operators on it.
There are at least two, quite different ways to do so. The first
uses the twisted direct product (the twisted Kronecker product)
of the EPR maps.
In the second one relies on ideas from representation
theory, and on an applications of
Tomita and Takesaki's theory. All the matter is quite elementary as
long as we are within type I factors.

\subsection{Twisted direct products}
The starting point for the following definition are two maps,
\begin{equation} \label{3.1}
\xi^{ba} \, : \, \cH_a \mapsto \cH_b, \quad
\eta^{ab} \, : \, \cH_b \mapsto \cH_a,
\end{equation}
both either linear or antilinear. The {\em twisted direct
product}, $\eta^{ab} \tilde \otimes \xi^{ba}$,
(with the {\em twisted cross $\tilde \otimes$}),
is defined by the linear or antilinear extension of
\begin{equation} \label{3.2}
\phi^a \otimes \phi^b \mapsto
(\eta^{ab} \tilde \otimes \xi^{ba})(\phi^a \otimes \phi^b)
:= \eta^{ab} \phi^b \otimes \xi^{ba} \phi^a
\end{equation}
The extension has to be linear if both factors are linear maps, and
antilinear if both maps are antilinear. Other cases,
one map linear and one antilinear, are ill defined. In the
admissible cases the Hermitian adjoint can be gained by
\begin{equation} \label{3.3}
(\eta^{ab} \tilde \otimes \xi^{ba})^* =
(\xi^{ba})^* \tilde \otimes (\eta^{ab})^*
\end{equation}
Useful is also
\begin{equation} \label{3.2a}
(\eta_1^{ab} \tilde \otimes \xi_1^{ba}) \circ
(\eta_2^{ab} \tilde \otimes \xi_2^{ba}) =
(\eta_1^{ab} \xi_2^{ba}) \otimes (\xi_1^{ba} \eta_2^{ab})
\end{equation}
Now let $\varphi, \psi \in \cH_a \otimes \cH_b$ an ordered
pair of vectors. Essentially, there are four twisted products
to perform:
\begin{eqnarray}
\tilde S_{\varphi, \psi} :=
{\bf j}_{\varphi} \tilde \otimes {\bf s}_{\psi}, \quad
\tilde F_{\varphi, \psi} :=
{\bf s}_{\varphi} \tilde \otimes {\bf j}_{\psi},
\label{twlift1}\\
\tilde \Delta_{\varphi, \psi} :=
{\bf s}_{\varphi} \tilde \otimes {\bf s}_{\psi}, \quad
J_{\varphi, \psi} :=
{\bf j}_{\varphi} \tilde \otimes {\bf j}_{\psi}
\label{twlift2}
\end{eqnarray}
The notations are ad hoc ones, with the exception of the last
(see below).
Because of (\ref{3.3}) the Hermitian adjoints of these operators
are gained by exchanging the roles of $\psi$ and $\varphi$.

We need the reduced density operators of $\psi$
and of $\varphi$. We call them $\omega^a_{\psi}$, ...,
$\omega^b_{\varphi}$. Their supporting
projections are denoted by $Q^a_{\psi}$, and so on.
To arrive at the polar decompositions we first notice
\begin{equation} \label{twlift3}
\tilde \Delta_{\psi, \varphi} \tilde \Delta_{\varphi, \psi} =
\omega^a_{\psi} \otimes \omega^b_{\varphi}, \quad
J_{\psi, \varphi} J_{\varphi, \psi} =
Q^a_{\psi} \otimes Q^b_{\varphi}
\end{equation}
Reminding the definition (\ref{3.2}) and the
polar decomposition of the EPR-maps one computes
the polar decompositions of the antilinear operators defined
above.
\begin{eqnarray}
\tilde \Delta_{\psi, \varphi} =
(\omega^a_{\psi} \otimes \omega^b_{\varphi})^{1/2} J_{\varphi, \psi} =
J_{\psi, \varphi} (\omega^a_{\varphi} \otimes \omega^b_{\psi})^{1/2}\\
\tilde S_{\psi, \varphi} =
(\omega^a_{\psi} \otimes Q^b_{\varphi})^{1/2} J_{\varphi, \psi} =
J_{\psi, \varphi} (Q^a_{\varphi} \otimes \omega^b_{\psi})^{1/2}
\nonumber\\
\tilde F_{\psi, \varphi} =
(Q^a_{\psi} \otimes \omega^b_{\varphi})^{1/2} J_{\varphi, \psi} =
J_{\psi, \varphi} (\omega^a_{\varphi} \otimes Q^b_{\psi})^{1/2}\\
\nonumber
\end{eqnarray}

\subsection{Contact with representation theory}
There is a representation of $\cB(\cH_a)$ with representation
space $\cH_a \otimes \cH_b$ associated with the embedding
$$
\cB(\cH_a) \mapsto \cB(\cH_a) \otimes \1^b \subset
\cB(\cH_a \otimes \cH_b)
$$
Assume that $\psi$ is a cyclic and separating vector, i.e. a
GNS-vector for the representation. Equivalently one requires
$Q^a_{\psi} = \1^a$ and $Q^b_{\psi} = \1^b$.
In the spirit of \cite{Schr35a}, one also calls $\psi$
{\em completely entangled}.

With a given second vector, $\varphi$, the
antilinear $S$ is defined by
\begin{equation} \label{tt1}
S_{\varphi, \psi} (A \otimes \1^b) \psi = (A^* \otimes \1^b)
\varphi
\end{equation}
for all $A \in \cH_a$. (\ref{tt1}) is a fundamental construct in
the theory of Tomita and Takesaki,
though, as we are concerned
with type I factors, an elementary one: In our case it is not
difficult to prove closability of $S$.
We denote the closure of $S$ again by $S$ and write the
polar decomposition in standard notation
\begin{equation} \label{tt2}
S_{\varphi, \psi} =
J_{\varphi, \psi} \Delta_{\varphi, \psi}^{1/2}, \quad
\Delta_{\varphi, \psi} = \omega^a_{\varphi} \otimes (\omega^b_{\psi})^{-1}
\end{equation}
see \cite{Haag93} for an introduction.
Having already defined $J$ in (\ref{twlift2}) as a twisted
Kronecker product, we have to show that it coincides
with the modular antiunitary operator defined in the theory
of Tomita and Takesaki for GNS-vectors $\psi$.
The most important case is the {\em modular conjugation}
$J_{\psi, \psi} \equiv J_{\psi}$. Remark that (\ref{twlift2})
is slightly more general than (\ref{tt1}): In the former
equation $\psi$ can be any vector in any
bipartite Hilbert space.

To prove the assertion we start with a decomposition of unity
$$
\1^a = \sum | \phi_k^a \rangle\langle \phi^a_k |
$$
to get, by the help of (\ref{10}), (\ref{11})
$$
(A \otimes \1^b) \psi =
\sum A \phi^A_k \otimes {\bf s}^{ba}_{\psi} \phi^a_k =
(1^a \otimes \sqrt{\omega^b_{\psi}}) (A \phi^a_k \otimes
{\bf j}^{ba}_{\psi} \phi^a_k),
$$
$$
(A^* \otimes \1^b) \varphi =
\sum  \phi^A_k \otimes {\bf s}^{ba}_{\varphi} A  \phi^a
= ({\bf j}_{\psi} \tilde \otimes {\bf j}_{\varphi})
(\sqrt{\omega^a_{\varphi} \otimes \1^b}) (A \phi^a_k \otimes
{\bf j}^{ba}_{\psi} \phi^a_k)
$$
and, finally,
\begin{equation} \label{tt3}
J_{\psi, \varphi} S_{\varphi, \psi}
(\sqrt{\omega^a_{\varphi} \otimes \1^b})
 = (1^a \otimes \sqrt{\omega^b_{\psi}})
\end{equation}
Because our starting assumption implies invertibility
of $\omega^a_{\psi}$, we may rewrite (\ref{tt3}) as asserted in
(\ref{tt1}).


email: armin.uhlmann@itp.uni-leipzig.de
\end{document}